\documentclass[aps,prl,showpacs,twocolumn]{revtex4}
\usepackage{graphics}
\usepackage{dcolumn}
\usepackage{bm}
\begin{document}
\title{ Behaviour of the potentials due to strangeness degree of freedom in 
$_{\Lambda\Lambda}^{\:\:\:6}$He hypernucleus}
\author{ A. A. Usmani$^{1,2}$}
\email{ anisul@iucaa.ernet.in}
\author{ Z. Hasan$^1$}
\affiliation
{$^1$Department of Physics, Aligarh Muslim University, Aligarh 202 002, India}
\affiliation{$^2$Inter University Centre for Astronomy and Astrophysics (IUCAA), 
Ganeshkhind, Pune-411 007, India}

\date{\today}
\begin{abstract}
Fully correlated study of $_{\Lambda\Lambda}^{\:\:\:6}$He hypernucleus 
has been performed with two- and three- baryon potentials. For the $S=-2$ 
sector, various simulations of Nijmegen $\Lambda\Lambda$ potential models
have been used. 
We investigate the role of every strength of the strange sector potentials 
on the energy breakdown and present a crystal clear understanding
of their interplay. Consistency of calculations
between $_\Lambda^5$He and $_{\Lambda\Lambda}^{\:\:\:6}$He depends 
on the  $\Lambda N$ space-exchange strength only.
Investigation limits  the strength of simulated Nijmegen 
$\Lambda\Lambda$ potentials. 
The study is a step forward to determine all these strengths, 
to resolve $A=5$ anomaly and to search for recently conjectured
$_{\Lambda\Lambda}^{\:\:\:4}$H in an authentic way. 

\end{abstract}


\maketitle


Strangeness degree of freedom, when trapped in a bound nuclear system,
affects every physical observable, starting from deuteron to neutron stars.
It induces subtle distortions in the system even with the presence of a single
hyperon~\cite{Gibson}. 
The hypernuclear sytems with one or two quanta of strangeness in a finite 
nucleus and also with the bulk of it as in an infinite-body hyperon star 
offer an unique opportunity to enrich our knowledge about the role of 
strangeness in a nuclear medium of different densities. Study of such systems
may provide useful informations about baryon-baryon and three-baryon forces. 
Notable advancements have been made on theoretical as well as on experimental 
frontiers of the subject. 

The recent unambiguous observation of six-baryon double-$\Lambda$ hypernucleus 
($_{\Lambda\Lambda}^{\:\:\:6}$He, NAGARA event) in the Japanese high energy 
laboratory (KEK) hybrid experiment~\cite{NAGARA} E373 and 
the evidence~\cite{Ahn01} for a bound four-baryon double-$\Lambda$ hypernucleus 
($_{\Lambda\Lambda}^{\:\:\:4}$H) as suggested by the Brookhaven 
alternating-gradient synchroton (BNL-AGS) experiment E906, have given fresh 
impetus to the  physics of $S=-2$ sector. Besides $_{\Lambda\Lambda}^{\:\:\:4}$H, 
we have three well-established double-$\Lambda$ hypernuclear species 
($_{\Lambda\Lambda}^{\:\:\:6}$He, $_{\Lambda\Lambda}^{10}$Be,
and $_{\Lambda\Lambda}^{12}$C)~\cite{NAGARA, Ahn01,double}, whose realistic study 
along with many single-$\Lambda$ hypernuclei having rich experimental statistics
with a wide range of baryon numbers and orbital angular momentum 
$\ell_\Lambda \le 3$ is a major thrust area of research.

A theoretical study requires a realistic Hamiltonian and a good wave 
function (WF) that includes all dynamical correlations induced 
by the potentials of the Hamiltonian. However, when the same is employed, 
computational complexities increase with increasing mass number $A$,
irrespective of the many-body technique involved. The Faddeev-Yakubovsky (FY) 
calculations~\cite{Nogga02} have not been extended to $A\ge 5$ baryon hypernuclei, 
because calculational dimensions expand to an unmanageable situation. However, in 
the cluster Faddeev-Yakubovsky (CFY) approach, calculations have been made 
possible to $A=5$ and 6 baryon hypernuclei~\cite{CFY, FG2Hiyama02}. 
Along with these, the fully coupled channel stochastic variational (FCCSV) 
approach~\cite{Nemura05} and the fully correlated variational Monte Carlo 
(FCVMC) study~\cite{AAU06,AAU2-06} have pushed ahead the frontiers of the 
subject. 
The $\Lambda N$ space exchange correlation ($SEC$) is built into all the above 
studies except for CFY. 
Being an important correlation, $SEC$ affects 
every physical observable in $_{\Lambda}^{5}$He~\cite{AAU06} and  
also in $_{\Lambda\Lambda}^{\:\:\:6}$He~\cite{AAU2-06}. 

A bound state for the recently conjectured  $_{\Lambda\Lambda}^{\:\:\:4}$H
is not found over a wide range of $\Lambda\Lambda$ strengths in the 
FY search~\cite{FG3-02}. But, a recent FCCSV search~\cite{NAM-03},
in contrast to it, predicts a bound  $_{\Lambda\Lambda}^{\:\:\:4}$H. 
Though, the search is not free of uncertainties.  
Various couplings of the 
$4\times4$ Hamiltonian matrix used as basic inputs in this approach are 
uncertain, specially those falling in the $S=-2$ sector for which 
there is no direct information from experiments in free space.
The many-body effects on these strengths in  a nuclear
medium constitute another important issue, which modify the free space results. 
The strong $\Lambda N - \Sigma N$ transition potential 
in the $S=-1$ sector also lacks complete preciseness.

Potentials related to these couplings have got strong tensorial 
dependence, therefore, are sensitive to other operators of the WF, 
specially to the tensor operator.  The expectation value
of tensor operator in a Jastrow WF for a closed-shell nucleus 
is zero, whereas expectation value for its square is nonzero. 
This may lead to  a quadratic dependence of the expectation value of the
potentials with respect to their strengths.  A slight variation in these 
may offset the variational WF and may result in an appreciable 
change in the energy breakdown and also in the  total ground-state energy.
Uncertainty in the strengths (basic input) leads to the uncertainty in the 
results as well as in the consistency of calculations. 

These factors should be addressed carefully.
It would be useful to investigate how results and consistency of the 
calculations are affected with the variations in the strengths and how do 
they interplay. 
We explore these important issues in this letter and investigate the role
of every potential strength that matter which is inevitable for precise 
determination of strengths, for the resolution of $A=5$ anomaly and
for an authentic search of $_{\Lambda\Lambda}^{\:\:\:4}$H, whose  
binding is a  subtle issue. 

We follow a different approach other than the coupled channel formalism. 
One may always project out $\Sigma$, $\Delta$, etc., degrees of freedom from 
this formalism. In the $S=-1$ sector, this would result in a three-baryon 
$\Lambda NN$ potential. It is written as a sum of two-pion exchange (TPE) 
attractive term and a repulsive term~\cite{BU88,Gal75,Bhaduri67} like its 
non-strange counter part, the $NNN$ potential. The repulsive term is 
suggested by the suppression mechanism due to $\Lambda N - \Sigma N$ 
coupling~\cite{UB99,BR71,Rozynek79,BR70}, which is a medium effect. 
For the $S=-2$ sector, we may use  simulated Nijmegen potential 
models~\cite{FG2Hiyama02}. Thus, our basic ingredients are two-baryon and 
three-baryon potentials. We then use a fully correlated WF as in 
Ref.~\cite{AAU06} written for all the s-shell single and double-$\Lambda$ 
hypernuclei. With such a Hamiltonian and WF, a microscopic study 
of six-baryon double-$\Lambda$ hypernucleus is easily manageable without 
loosing any essential physics. As an advantage, it may also be applied to the 
closed-shell $_{\:\:\Lambda}^{17}$O hypernucleus~\cite{UPU95} in the framework 
of cluster Monte Carlo technique~\cite{PWP92}.

Variations in the strengths directly affect 
the expectation values of the respective potentials and also the WF as 
correlations are nothing but the solutions of these potentials. 
Moreover, there are 
sensitivities among various terms of the Hamiltonian and the operators  of the
WF. Thus, a change in any of the  strengths may affect the complete energy 
breakdown. Bearing these factors in mind, we proceed in a systematic way. 
We perform a FCVMC study~\cite{AAU06} of $_{\Lambda}^{5}$He hypernucleus, 
using a realistic Hamiltonian and a fully correlated WF. 
It suggests that  a study ignoring $SEC$ would be misleading. 
In a subsequent study, we calculate $\Lambda$-separation 
energy ($B_{\Lambda}=E_{^4{\rm He}}-E_{_{\Lambda}^{5}{\rm He}}$) and obtain 
solutions of all the strengths to reproduce  experimental  
$B_{\Lambda}^{exp}$=3.12(2) MeV for the range of the strengths~\cite{UK06}.  
In order to know how do behaviours change with two quanta of strangeness, 
we extend our study to  $_{\Lambda\Lambda}^{\:\:\:6}$He 
hypernucleus~\cite{AAU2-06},  where $SEC$ effects are found  more evident 
because of the presence of a pair of $\Lambda$ hyperons. 
We then aim to study the behaviour of all the strengths for this hypernucleus 
herein.

For the $S=-2$ sector, we use various Nijmegen models representing 
$^1S_0$ $\Lambda\Lambda$ potential, which  are simulated to a phase equivalent 
three-range Gaussian form~\cite{FG2Hiyama02,Rijken,Hiyama97} 
\begin{eqnarray}
\label{vLL}
v_{\Lambda\Lambda}(r)&=& 
9324 \: {\rm exp}\left(-\frac{r^2}{0.35^2}\right)
-379.1 \:\gamma \:  {\rm exp}\left(-\frac{r^2}{0.777^2}\right) \nonumber \\
&&-21.49 \: {\rm exp}\left(-\frac{r^2}{1.342^2}\right).
\end{eqnarray}
The dimensionless quantity $\gamma$ distinguishes amongst various Nijmegen 
potential models. For example,  NSC97e, ND, and NEC00, which are represented 
by $\gamma$=0.5463, $ \gamma$=1.0 and $\gamma$=1.2044, 
respectively.  The $\Lambda\Lambda$-separation energy, 
($B_{\Lambda\Lambda}=E_{^4{\rm He}}-E_{_{\Lambda\Lambda}^{\:\:\:6}{\rm He}}$)
therefore, depends on the choice of the potential, which should be taken with 
caution. The expectation value of the $\Lambda\Lambda$ potential would 
depend upon the choice of $\gamma$.  It may affect the WF 
through its self induced correlation due to same reason mentioned above.
For the $S=-1$ sector, we use charge symmetric $\Lambda N$ potential~\cite{BUC84} 
\begin{equation}
\label{vLN}
v_{\Lambda N}(r)=[v_c(r)-\overline{v}T_{\pi}^{2}(r) ] 
(1-\varepsilon+\varepsilon P_{x}) +\frac{v_{\sigma}}{4}T_{\pi}^{2}(r)
\mbox{\boldmath$\sigma$}_{\Lambda}\cdot\mbox{\boldmath$\sigma$}_{N}.
\end{equation}
Here, $\varepsilon$ determines the odd-state potential, which is the 
strength of the space-exchange potential relative to the direct potential.  
$ v_c(r)$ is the Saxon-Woods core and $T_\pi(r)$ is the one-pion tensor shape
factor.  The constants, $\overline{v}$ and $v_{\sigma}$, are 
respectively the spin-average and spin-dependent strengths. 

The  Hamiltonian for the $A$-baryon double-$\Lambda$ hypernucleus reads as
\begin{equation}
H=H_{NC}+H_{\Lambda_1}+H_{\Lambda_2}+v_{\Lambda_1\Lambda_2}
\end{equation}
\begin{equation}
H_{NC}=T_{NC} +\sum_{i< j}^{A-2} v_{ij} +\sum_{i< j < k}^{A-2} V_{ijk}, 
\end{equation}
\begin{equation}
H_{\Lambda_n}=T_{\Lambda_n}+\sum_i^{A-2} v_{\Lambda_n i}+\sum_{i< j}^{A-2} 
V_{\Lambda_n ij}.
\end{equation}
Here, $H_{NC}$ is the  nuclear core ($NC$) Hamiltonian and $H_{\Lambda_n}$ is 
the Hamiltonian arising due to an individual  $\Lambda_n$.
Subscripts $i,j$ and $k$ refer to nucleons. Obviously,
$H_{NC}+H_{\Lambda_n}$ is  the Hamiltonian for the  $A-1$ baryon 
single-$\Lambda$ hypernucleus.  For the $S=0$ sector, we use well established 
Argonne $v_{18}$~\cite{AV18} $NN$ potential and Urbana type $NNN$ 
potential~\cite{NNN-IX-NNN}. The $\Lambda NN$ potential is the sum of a repulsive
dispersive term~\cite{BU88} 
\begin{equation}
\label{VLNND}
V_{\Lambda ij}^{D}=
W^{D}T_{\pi}^{2}(r_{\Lambda i})T_{\pi}^{2}(r_{\Lambda j})
  [1+\mbox{\boldmath$\sigma$}_{\Lambda}\cdot(
\mbox{\boldmath$\sigma$} _{i}+ \mbox{\boldmath$\sigma$}_{j})/6]
\end{equation}
and a $TPE$ term for P- and S-wave $\pi-N$ scatterings~\cite{Bhaduri67} 
\begin{equation}
\label{VP}
V^{P}_{\Lambda ij}=-\left(C^{P}/6\right)
(\mbox{\boldmath$\tau$}_{i}\cdot
\mbox{\boldmath$\tau$}_{j})
\{X_{i\Lambda},X_{\Lambda j}\},
\end{equation}
and
\begin{equation}
\label{VS}
V^{S}_{\Lambda ij}=C^{S}
Z(  r_{i\Lambda})Z( r_{j\Lambda })
\mbox{\boldmath$\sigma$}_i\cdot {\hat{\bf r}}_{i\Lambda }
\mbox{\boldmath$\sigma$}_j\cdot {\hat{\bf r}}_{j\Lambda}
\mbox{\boldmath$\tau$}_i\cdot\mbox{\boldmath$\tau$}_j 
\end{equation} 
with
\begin{equation}
\label{XiL}
X_{\Lambda i}=( \mbox{\boldmath$\sigma$}_{\Lambda}\cdot  
\mbox{\boldmath$\sigma$}_{i} )Y_{\pi}(r_{\Lambda i})+S_{\Lambda i}
T_{\pi}(r_{\Lambda i})
\end{equation} 
and
\begin{equation}
Z(r)=\frac{m_\pi r}{3} [Y_\pi(r)-T_\pi(r)].
\end{equation}
Here, $W^D$, $C^P$ and $C^S$ are the strengths and $Y_\pi(r)$
 is the Yukawa function.

\begin{figure}
\includegraphics{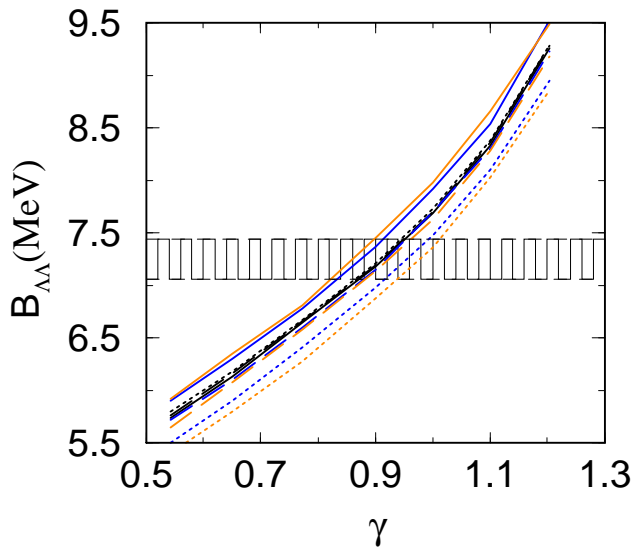}
\caption{\label{figone}
The dotted, dashed and long dashed curves represent three values of
$\varepsilon$; 0.1, 0.2 and 0.3, respectively. 
Each of them with black, blue and orange colors represent 
$\overline{v}$=6.15, 6.10 and 6.05 MeV, respectively. 
Curves for $\overline{v}$=6.15 MeV with different $\varepsilon$ almost
overlap throughout the range of $\gamma$ due to 
$\partial B_{\Lambda\Lambda}/\partial\varepsilon\approx$0.
The shaded area shows $B_{\Lambda\Lambda}^{exp}$ with 
unceertainty.
 }
\end{figure}
 \begin{table}
\caption{\label{tab1}Energy breakdown of the $_{\Lambda\Lambda}^{\:\:\:6}$He 
hypernucleus for $\varepsilon$=0.2 and $\overline{v}$=6.10 MeV. 
All quantities are in units of MeV.
Subsripts $i,j$ and $k$ refer to nucleons, and $\lambda$ to $\Lambda$
hyperons.
}
\begin{ruledtabular}
\begin{tabular}{lccc}
 &    $C^P=$0.75  &  $C^P=$1.5  & A-B \\   
& A   & B  &        \\
\hline
 $T_{\Lambda}=T_{\Lambda_1}+ T_{\Lambda_2}$        
&22.94(6)   & 22.64(6)  & 0.30(8)       \\
 $v_0(r)(1-\varepsilon)$ 
& -29.64(8) &-28.01(7) & -1.63(11)   \\
$v_0(r)\varepsilon P_x$ 
& -6.46(2)  &-5.96(2)    &   0.50(3)\\
 $(\frac{1}{4})v_\sigma T_\pi^2(r) \mbox{\boldmath$\sigma$}_\lambda\cdot\mbox{
\boldmath$\sigma$}_i$ 
&  0.025(0)   & 0.061(0)  & -0.036(0)\\
 $v_{\lambda i}$  
&-36.07(9)  &-33.91(9)    &-2.16(13) \\
 $v_{\Lambda\Lambda}  = v_{\Lambda_1 \Lambda_2} \:\:\:\:(\gamma=1.0)$  
&-5.74(5)   &-5.44(4)     & -0.30(6)  \\
 $V_{\lambda ij} ^D$   
& 5.03(2)  & 11.37(4)     & -6.34(4)  \\
 $V_{\lambda ij}^{ P}$ 
& -6.00(2) & -16.34(6)    &  10.34(6) \\
 $V_{\lambda ij}^{ S}$   
& -0.015(0) & 0.075(2)    & -0.090(2)\\
 $V_{\lambda ij}^{2\pi}$ $= V_{\lambda ij}^{P} + V_{\lambda ij}^{ S}$ 
& -6.01(2) & -16.27(5)    & 10.26(5) \\
 $V_{\lambda ij}=V_{\lambda ij}^D +V_{\lambda ij}^{2\pi}$ 
& -0.99(2)   & -4.90(3)   & 3.91(3) \\
 $V_\Lambda{}=v_{\lambda i}+v_{\lambda \lambda}+V_{\lambda ij}$ 
&  -42.80(8)  & -44.25(9)   & 1.45(12) \\
 $E_\Lambda=T_{\Lambda}+ V_\Lambda^{}$
& -19.86(5)  & -21.61(6)   & 1.75(8) \\
 $T_{NC}$     
&  121.31(16)  & 122.15(16)  & -0.84(23) \\
 $v_{ij}$     
& -131.41(15)  & -130.23(14)    & -1.18(21) \\
 $V_{ijk}$    
& -5.52(2)   & -5.74(2)    & 0.22(3) \\
 $V_{NC}=v_{ij}+V_{ijk}$  
& -136.93(15)  & -135.97(15)   & -0.96(21) \\
 $E_{NC}=T_{NC}+V_{NC}$  
& -15.61(7)  &-13.82(6)    & -1.79(9) \\
$E_{_{\Lambda\Lambda}^{\:\:\:6}{\rm He}}=E_\Lambda +E_{NC}$ 
& -35.47(2)  &-35.43(3)    & -0.04(3)\\
 $B_{\Lambda\Lambda}$  
 & 7.74(4) & 7.70(4)    & 0.04(6) \\
\end{tabular}
\end{ruledtabular}
\end{table}

In order to know the role of strange sector potential strengths and the 
various sensitivities among them, we must analyse our present findings in the 
light of the findings of previous studies~\cite{AAU06,AAU2-06,UPU95,AAU03}.
A couple of simplifications arise; 
(i) due to strong suppression of S-wave $\Lambda NN$ potential as its 
non-strange counterpart S-wave $NNN$ potential~\cite{Bhaduri67} 
(ii) and also due to  weak spin part of the $\Lambda N$ potential for the 
spin-zero core nucleus.  Therefore, variations in the strengths of these 
potentials, $C^S$ and $v_\sigma$, 
are hardly noteworthy. The $C^S$ is, however, reasonably fixed to 1.5 MeV 
as in Ref.~\cite{UK06}.  Therefore, strengths 
which matter are  $\overline{v}$, $\varepsilon$, $C^P$ and $W^D$.

The TPE potential has a generalised tensor type structure. It 
is sensitive to operators and specially to tensor operator, 
hence to its self induced correlation as mentioned before with reason. 
Variation in its strength, $C^P$, offsets the WF, hence complete 
energy breakdown. On the other hand, variation in $\varepsilon$  affects 
the baryon density profiles through the repulsive central $\Lambda N$ 
correlations and $SEC$~\cite{AAU06} as they are solutions of the 
Schr\"{o}dinger equation involving $\varepsilon$ in the $\Lambda N$ potential. 
The  $\Lambda N$ central correlation pushes the 
nucleons at the centre and towards the periphery, however, $SEC$ weakens this 
effect~\cite{AAU06,AAU2-06}.  This leads to modifications in densities 
with changing  $\varepsilon$, which affects the complete 
energy breakdown, even its central pieces. The density effect 
directly appears in the TPE potential through 
$T_\pi(r)$ and $Y_\pi(r)$ functions, 
which in turn affects the WF through its sensitivity with operators. 
Thus,  $C^P$ is correlated with  $\varepsilon$. 

The WF remains unaffected with the variations in $W^D$. Therefore, a linear 
relationship between $W^D$ and the expectation value of its potential is 
observed. 
A change in density profiles due to variations in $\varepsilon$ affects
the expectation value, hence the slope of this relationship.
The $v_{\Lambda i}$, for spin-zero core nucleus, is almost a central quantity 
due to weak spin part.  Its expectation value depends on the strength 
$\overline{v}$, $\varepsilon$, the average value $\langle P_x\rangle$
and baryon  density profiles. The $\langle P_x\rangle$ is always found
less than one about 0.87. Hence $v_{\Lambda i}$ (being a negative quantity)
decreases with decreasing $\varepsilon$. The $v_{\Lambda\Lambda}$ falls 
in the same line. The $V_{\Lambda ij}$, which is the sum of TPE and dispersive 
potentials, has got an opposite trend. This with the $NC$ part of the energy 
($E_{NC}$) resists any change in $v_{\Lambda i}$ and $v_{\Lambda\Lambda}$ 
caused due to variation in $\varepsilon$. 

A linear relationship between $W^D$ and $\varepsilon$ is found to be existing
for both the hypernuclei: $_\Lambda^5$He~\cite{AAU06} and  
$_{\Lambda\Lambda}^{\:\:\:6}$He~\cite{AAU2-06}, but with a different slope.
For the $_\Lambda^5$He, the relationship has been observed for a range of 
$C^P$~\cite{UK06}.  Thus, we may always find a suitable $W^D$ to repoduce 
$B_\Lambda^{exp}$ against any change in $\varepsilon$.  
As expected, results vary with the variations in $\overline{v}$ 
at any value of $C^P$. The above mentioned slope  is significantly affected 
against any change in  $\overline{v}$ in case of
$_{\Lambda\Lambda}^{\:\:\:6}$He, which was of lesser importance in case of 
 $_\Lambda^5$He. 
As a result, different combinations of $\varepsilon$ 
and $W^D$ that reproduce $B_\Lambda^{exp}$, do not converge at the same 
$B_{\Lambda\Lambda}$ except for $\overline{v}$=6.15 MeV. At this value of 
$\overline{v}$, a change in $v_{\lambda i}$ and $v_{\Lambda\Lambda}$ is 
balanced by an opposite effect in $V_{\Lambda ij}$ and $E_{NC}$ giving 
$\partial B_{\Lambda\Lambda}/\partial\varepsilon\approx$0., which 
is an accident. With decreasing $\overline{v}$ we notice an increase  
in $\partial B_{\Lambda\Lambda}/\partial\varepsilon$, because changes in 
$V_{\Lambda ij}$ and $E_{NC}$ win over the changes in $v_{\lambda i}$ 
and $v_{\Lambda\Lambda}$. 

We first reproduce $B_{\Lambda}^{exp}(\overline{v}, \varepsilon,C^P,W^D)$ for 
the range of the strengths. Thereafter, using the same strengths, we plot a 
sensitivity
graph between $B_{\Lambda\Lambda}(\overline{v}, \varepsilon,C^P,W^D,\gamma)$ 
and the only free strength $\gamma$.  This we report in Fig.~\ref{figone}. 
The $B_{\Lambda\Lambda}$ increases with increasing value of $\gamma$. However,
all the above mentioned features remain the same. We notice an important result 
at $C^P$=0.75 MeV that the sets of strengths even with different $\overline{v}$,  
which reproduce $B_{\Lambda}^{exp}$, yield the same value of 
$B_{\Lambda\Lambda}$ at $\varepsilon\approx$0.18 irrespective of the value 
of $\gamma$. Thus, by adjusting  $\gamma$, we may reproduce 
$B_{\Lambda\Lambda}^{exp}$=7.25(19) MeV for the same strengths that reproduce  
$B_{\Lambda}^{exp}$.  
Therefore, $\varepsilon\approx$0.18 turns out to be the 
only condition of consistency. 
Millener~\cite{Millener} too has suggested a small value for $\varepsilon$. 
We observe  that $0.85<\gamma<0.95$
reproduces $B_{\Lambda\Lambda}^{exp}$ at this value of $\varepsilon$. 

As variations in $C^P$ leads to an appreciable change in the above mentioned 
slope ~\cite{UK06} and in the energy breakdown, we double the value of 
$C^P$ from 0.75 to 1.5 MeV and repeat our calculations with a suitable $W^D$
that reproduces $B_\Lambda^{exp}$. 
Using same strengths, we perform calculations for  $B_{\Lambda\Lambda}$.  
Results for both the  values of $C^P$ are presented in Table~\ref{tab1} for 
$\overline{v}$=6.10 MeV and $\varepsilon$=0.2.  This value of $\varepsilon$
is very close to the value of consistency ($\varepsilon\approx$0.18).  
Although, we observe a significant effect in every energy piece
as $C^P$ offsets the WF,  we find close predictions 
for $B_{\Lambda\Lambda}$ for both the values of $C^P$.
The total energy remains almost the same 
because of the change in $NC$ part of the energy is 
balanced by an opposite change in $\Lambda$ part of the energy.
Therefore, condition of consistency remains unaltered irrespective of the 
value of $C^P$.

We notice more than three fold increase in the expectation 
value of TPE potential corresponding to two fold increase in its strength 
which suggests a quadratic behaviour.
Strong  effect in the $NC$ part of the energy ($E_{NC}$) is observed. As a 
result, $NC$ is more polarised for higher value of $C^P$.  
Nemura et al.~\cite{Nemura02} too have
noticed strong sensitivity between $E_{NC}$ and the tensorial 
$\Lambda N - \Sigma N$ transition potential. 

A similar study of $_{\Lambda}^{4}$H,  $_{\Lambda}^{4}$H$^*$ and  
$_{\Lambda\Lambda}^{\:\:\:5}$H hypernuclei may decide the strengths 
of strange sector potentials including $\gamma$ in a single shot, and hence 
may resolve the  $A=5$ anomaly~\cite{Gal75,DHT72} with no additional effort.
This investigation is under consideration, which would  be followed by 
an authentic  search  of $_{\Lambda\Lambda}^{\:\:\:4}$H. 
The knowledge would be helpful to bridge the 
gap in our fundamental understanding of baryon-baryon forces. 

Some improvements in the Hamiltoinian may be suggested.  
One may also adopt Green's function Monte Carlo method. 
But these would be mere refinements. 

The work was supported under Grant No. SP/S2/K-32/99 awarded to AAU 
by the Department of Science and Technology, Government of India.


\begin{thebibliography} {299}
\bibitem{Gibson} B. F. Gibson and  E. V. Hungerford III, 
                   Phys. Rep. {\bf 257}, 349 (1995).
\bibitem{NAGARA} H. Takahashi {\it et al.}, Phys. Rev. Lett {\bf 87}, 
                 121502 (2001).
\bibitem{Ahn01} K. Ahn {\it et al.}, Phys. Rev. Lett {\bf 87}, 
                 132504 (2001).
\bibitem{double} M. Danysz {\it et al.}, Nucl. Phys. {\bf 49}, 121 (1963);
                   Phys. rev. Lett. {\bf 11}, 29 (1963); 
                 R. H. Dalitz, D. H. Davis, P. H. Fowler, A. Montwill, 
                   J. Pneiswki and J. A. Zakrzewski, Proc. R. Soc. London, 
                   Ser. A {\bf 426}, 1 (1989);
                 S. Aoki  {\it et al.}, Prog. Theor. Phys.  {\bf 85}, 
                   1287 (1991);
                 C. B. Dover, D. J. Millener, A. Gal and D. H. Davis, 
                  Phys. Rev. C {\bf 44}, 1905 (1991).
\bibitem{Nogga02} A. Nogga, H. Kamada and W. Glockle,
                  Phys. Rev. Lett. {\bf 88}, 172501(2002).
\bibitem{CFY} I. N. Filikhin, A. Gal and V. M. Suslov, Phys. Rev. C {\bf 68}, 
               024002 (2003); I. N. Filikhin and A. Gal, Nucl. Phys. 
               {\bf A707}, 491 (2002).
\bibitem{FG2Hiyama02} I. N. Filikhin and  A. Gal, Phys. Rev. C
                  {\bf 65}, 041001(R) (2002);
                 E. Hiyama, M. Kamimura, T. Motoba, T. Yamada and 
                   Y. Yamamoto, Phys. Rev. Lett. {\bf 89}, 142508 (2002);
                   Phys. Rev. {\bf C66}, 024007(2002).
\bibitem{Nemura05} H. Nemura, S. Shinmura, Y. Akaishi and K. S. Myint,
                   Phys. Rev. Lett. {\bf 94}, 202502 (2005).
\bibitem{AAU06} A. A. Usmani, Phys. Rev. C {\bf 73}, 011302(R) (2006).
\bibitem{AAU2-06} A. A. Usmani and Z. Hasan, Phys. Rev. C {\bf 74}, 034320 (2006).
\bibitem{FG3-02} I. N. Filikhin and  A. Gal, Phys. Rev. Lett. 
                  {\bf 89}, 172502 (2002).
\bibitem{NAM-03} H. Nemura, Y. Akaishi and K. S. Myint, 
                 Phys. Rev. C {\bf 67}, 051001(R) (2003).
\bibitem{BU88} A. R. Bodmer and Q. N. Usmani,
                 Nucl. Phys. {\bf A477}, 621 (1988).
\bibitem{Gal75} A. Gal, Adv. Nucl. Phys. {\bf 8}, 1 (1975).
\bibitem{Bhaduri67} R. K. Bhaduri, B. A. Loiseau and Y. Nogami,
                 Anns. Phys. (N. Y) {\bf 44}, 57 (1967).
\bibitem{UB99} Q.N. Usmani and  A.R. Bodmer, 
                  Phys. Rev. {\bf C60}, 055215 (1999).
\bibitem{BR71} A.R. Bodmer and D.M. Rote Nucl. Phys. {\bf A169}, 1 (1971).
\bibitem{Rozynek79} J. Rozynek and J. Dabrowski, 
                  Phys. Rev. {\bf C20}, 1612 (1979); 
                   J. Dabrowski and  J. Rozynek,
                  {\it ibid}. {\bf 23}, 1706(1981);
                  Y. Yamamoto and H. Bando, Suppl. Prog. Theor. Phys. Suppl. 
                {\bf 81}, 9(1985); Y. Yamamoto, Nucl. Phys. {\bf A450}, 275c 
                  (1986). 
\bibitem{BR70} A.R. Bodmer, D.M. Rote and A.L. Mazza, 
                  Phys. Rev. {\bf C2}, 1623 (1970). 
\bibitem{UPU95} A. A. Usmani, S. C. Pieper and Q. N. Usmani,
                 Phys. Rev. {\bf C51}, 2347 (1995).
\bibitem{PWP92} S. C. Pieper, R. B. Wiringa and V. R. Pandharipande,
                 Phys. Rev. {\bf C46}, 1741 (1992).
\bibitem{UK06} A. A. Usmani and. F. C. Khanna, submitted to Phys. Rev. C.
\bibitem{Rijken}   Th. A. Rijken {\it et al.}, Phys. Rev C{\bf 59}, 21 (1999).
\bibitem{Hiyama97} E. Hiyama, M. Kamimura, T. Motoba, I. Yamada and I. Yamamoto,
                   Prog. Theor. Phys. {\bf 97}, 881 (1997). 
\bibitem{BUC84} A. R. Bodmer and Q. N. Usmani and J. Carlson, 
                  Phys. Rev. {\bf C29}, 684 (1984);
                 I. E. Lagaris and V. R. Pandharipande,
                 Nucl. Phys. {\bf A359}, 331 (1981).
\bibitem{AV18} R. B. Wiringa, V. G. J. Stoks, and R. Schiavilla, Phys.
              Rev. {\bf C51}, 38 (1995).
\bibitem{NNN-IX-NNN} B. S. Pudliner, V. R. Pandharipande, J. Carlson
                and R. B. Wiringa, Phys. Rev. Lett. {\bf 74}, 4396 (1995);
         J. Carlson,  V. R. Pandharipande and R. B. Wiringa, 
               Nucl. Phys.  {\bf A401}, 59 (1983).
\bibitem{AAU03} A.A. Usmani and S. Murtaza, Phys. Rev. {\bf C68}, 024001 (2003); 
                    A. A. Usmani, Phys. Rev. {\bf C52}, 1773 (1995).
\bibitem{Millener} D. J. Millener,
                   Nucl. Phys. {\bf A691}, 93c (2001).
\bibitem{Nemura02} H. Nemura, Y. Akaishi and Y. Suzuki,
                   Phys. Rev. Lett. {\bf 89}, 142504 (2002).
\bibitem{DHT72} R. H. Dalitz, R. C. Herndon, and Y. C. Tang,
                 Nucl. Phys. {\bf B47}, 109 (1972);
                  E. V. Hungerford and L. C. Biedenhorn, Phys. Lett. 
                   {\bf 142B}, 232 (1984).
\end{thebibliography}
\end{document}